\documentstyle[twocolumn,graphicx,aps]{revtex}

\begin{document}
\draft

\title{Correlation functions and momentum distribution of one-dimensional
Bose systems}

\author{G.E. Astrakharchik and S. Giorgini}

\address{
{\small\it  $^{1}$Dipartimento di Fisica, Universit\`a di Trento, and BEC-INFM, 
I-38050 Povo, Italy}
\\ (\today)
\\ \medskip}\author{\small\parbox{14.2cm}{\small\hspace*{3mm}
The ground-state correlation properties of a one-dimensional Bose system 
described by the Lieb-Liniger Hamiltonian are investigated by using exact 
quantum Monte Carlo techniques. The pair distribution function, static 
structure factor, one-body density matrix and momentum distribution of 
a homogeneous system are calculated for different values of the gas parameter 
ranging from the Tonks-Girardeau to the mean-field regime. Results for the 
momentum distribution of a harmonically trapped gas in configurations 
relevant to experiments are also presented. \\
\\[3pt]PACS numbers: 03.75.Fi, 05.30.Fk, 67.40.Db}} \maketitle

\narrowtext

The recent experimental achievements in realizing quasi one-dimensional (1D)
quantum degenerate Bose gases confined in harmonic traps \cite{MIT,Paris,Munich}, 
have attracted a great interest on the ground-state properties of these systems. 
The 1D regime is reached in highly anisotropic traps, where the axial motion of
the atoms is weakly confined while the radial motion is frozen to zero 
point oscillations by the tight transverse confinement. In these conditions and if 
the characteristic range of interparticle interactions is much smaller than the 
typical length of the transverse confinement, the system can be  
described by the Lieb-Liniger model of $\delta$-interacting 1D bosons.
For a homogeneous system the Hamiltonian has the form
\begin{equation}
H_{LL}= - \frac{\hbar^2}{2m}\sum_{i=1}^N\frac{\partial^2}{\partial z_i^2}
+g_{1D}\sum_{i<j}\delta(z_i-z_j) \;,
\label{HLL}
\end{equation} 
where $m$ is the atomic mass and $g_{1D}=2\hbar^2/m |a_{1D}|$ is the coupling
constant in terms of the 1D effective scattering length $a_{1D}$ 
\cite{Olshanii}. Recent numerical simulations using exact quantum Monte Carlo 
techniques have explicitly shown the cross-over from the three-dimensional (3D) 
to the 1D Lieb-Liniger regime for the energetics and structural properties 
of trapped systems \cite{Blume,US}. The Hamiltonian $H_{LL}$ is expected to provide 
the correct description also for the correlation properties of a quasi 1D 
system. In this Letter we calculate for the first time, using exact quantum Monte 
Carlo methods, the pair distribution function and the one-body density matrix of
a Lieb-Liniger gas in its ground state. By taking the Fourier transform of the 
one- and two-body correlation function we obtain results for the momentum 
distribution and the static structure factor of the system respectively. 
These correlation functions can be accessed in experiments using  
ballistic expansion and Bragg spectroscopy techniques and can provide a signature 
of the exotic properties of these quantum degenerate 1D systems.

The equation of state and excitation spectrum of a homogeneous system of bosons 
described by the Hamiltonian (\ref{HLL}) have been calculated exactly in the 
thermodynamic limit in Ref. \cite{LL} for any value of the dimensionless gas 
parameter $n|a_{1D}|$, where $n$ is the particle density. If $n|a_{1D}|\gg 1$, 
the system is in the weakly interacting regime and mean-field theory can be 
applied. In this case, the energy per particle is given by the Gross-Pitaevskii 
(GP) prediction: $E/N = g_{1D}n/2$. In the opposite limit, $n|a_{1D}|\ll 1$, the 
system enters the Tonks-Girardeau (TG) regime of a gas of impenetrable bosons. 
In this regime, which has been first investigated by Girardeau \cite{Girardeau}, 
the system acquires fermionic properties in the sense that there exists an exact 
mapping between the wave function of the interacting bosons and the wave function 
of a non-interacting Fermi gas. The energy per particle coincides in this case 
with the Fermi energy: $E/N=\pi^2\hbar^2n^2/(6m)$.

We study the following correlation functions: the one-body density matrix,
which in terms of the ground-state wave function $\Psi_0(z_1,..,z_N)$ is 
defined as
\begin{equation}
g_1(z)=\frac{N}{n}\frac{\int \Psi_0^\ast(z_1+z,..,z_N)\Psi_0(z_1,..,z_N)
\; dz_2 . . dz_N}{\int |\Psi_0(z_1,..,z_N)|^2 \;
dz_1 . . dz_N} \;,
\label{OBDM}
\end{equation}
and the pair distribution function
\begin{equation}
g_2(|z_1-z_2|)=\frac{N(N-1)}{n^2}
\frac{\int |\Psi_0(z_1,..,z_N)|^2 \; dz_3 . . dz_N}
{\int |\Psi_0(z_1,..,z_N)|^2 \; dz_1 . . dz_N} \;.
\label{pair}
\end{equation}
By taking the Fourier transform of these correlation functions one obtains
the momentum distribution $n(k)=n \int dz\; e^{ikz} g_1(z)$ and the static
structure factor $S(k)=1+n \int dz \; e^{ikz} [g_2(z)-1]$.

For a system of impenetrable bosons ($n|a_{1D}|\ll 1$) the above correlation 
functions can be calculated analytically. By exploiting the Bose-Fermi 
mapping one finds the following result for the pair distribution function
\cite{Girardeau}
\begin{equation}
g_2(z)=1-j_0^2(\pi n|z|) \;,
\label{g2tonks}
\end{equation}
where $j_0$ is a spherical Bessel function. The corresponding static 
structure factor is given by 
\begin{equation}
S(k)=\left\{  \begin{array}{cc} |k|/(2\pi n) & (|k|<2\pi n)   \\
                                   1         & (|k|>2\pi n)  \;. 
\end{array} \right. 
\label{sktonks}
\end{equation}
The one-body density matrix of a TG gas has also been 
calculated by performing analytic expansions for short and long 
distances (see Ref. \cite{OLS-DUN} and references therein). The 
long-range behavior is given by $g_1(z)\propto 1/\sqrt{|z|}$ 
\cite{Lenard}, corresponding to an infrared divergence in the 
momentum distribution $n(k)\propto 1/\sqrt{|k|}$.  

Beyond the TG regime full closed-form expressions of 
the correlation functions are not known. The long-range asymptotics
of the one-body correlation function can be obtained from the 
hydrodynamic theory of low-energy excitations \cite{Reatto,Schwartz}. 
One finds the following power-law decay
\begin{equation}
g_1(z)\propto 1/|z|^\alpha \;,
\label{longOBDM}
\end{equation}
where $\alpha=mc/(2\pi\hbar n)$ is fixed by the density $n$ and by the 
speed of sound $c$. The above result holds for $|z|\gg\xi$ where 
$\xi=\hbar/(\sqrt{2}mc)$ is the healing length of the system. The 
velocity of sound can be obtained from the equation of state through the 
compressibility of the system $mc^2=n\partial\mu/\partial n$, where 
$\mu=dE/dN$ is the chemical potential. In the TG regime 
($n|a_{1D}|\ll 1$) one finds $mc=\pi\hbar n$ and thus $\alpha=1/2$ as 
anticipated above. In the opposite GP regime ($n|a_{1D}|\gg 1$),
the result is $\alpha=1/(\pi\sqrt{2n|a_{1D}|})$ which
decreases as $n|a_{1D}|$ increases. Of course, result (\ref{longOBDM}) 
excludes the existence of Bose-Einstein condensation even in the 
ground-state \cite{Schultz}. The infrared behavior of the 
momentum distribution follows immediately from Eq. (\ref{longOBDM})
\begin{equation}
n(k)\propto 1/|k|^{1-\alpha} \;,
\label{nofkinfrared}
\end{equation}
holding for $|k|\ll 1/\xi$.  Furthermore, the hydrodynamic theory allows one to 
calculate the static structure factor in the long-wavelength regime
$|k|\ll 1/\xi$. One finds the well-known result
\begin{equation}
S(k)\to \hbar |k|/(2mc) \;,
\label{sofkinfrared}
\end{equation}
characteristic of phonon excitations. Recently, the short range behavior 
of the one- and two-body correlation functions has also been investigated. 
The value at 
$z=0$ of the pair correlation function can be obtained from the equation 
of state through the Hellmann-Feynman theorem \cite{Gangardt}: $g_2(z=0)=
-(m|a_{1D}|/\hbar^2)n^2d[(E/N)/n^2]/dn$. The result approaches zero in the 
TG regime 
and unity in the GP regime. Furthermore, the first few terms of the 
short-range series expansion of the one-body correlation function 
$g_1(z)=1+\sum_{i=1}^{\infty}c_i|nz|^i$ have been calculated in Ref. 
\cite{OLS-DUN} again exploiting the knowledge of the equation of state.

In this Letter we provide a complete calculation of the spatial 
dependence of the one- and two-body correlation functions for values of the 
gas parameter ranging from the TG to the deep GP regime. The calculation is 
carried out using the diffusion Monte Carlo (DMC) method, which allows one 
to solve exactly, apart from statistical uncertainty, the many-body 
Schr\"odinger equation for the ground state of a Bose system \cite{BORO1}. 
We consider a system of $N$ particles described by the Hamiltonian (\ref{HLL}), 
with periodic boundary conditions. The system has size $L$ and linear 
density $n=N/L$. Importance sampling is used through the trial wave function 
$\psi_T(z_1,..,z_N)=\prod_{i<j}f(z_{ij})$, where $f(z)$ is a two-body term 
which we choose as 
\begin{equation}
f(z)=\left\{  \begin{array}{cc} A\cos[k(|z|-B)]     & (|z|<\bar{Z})  \\
                                \sin^\beta(\pi|z|/L) & (|z|>\bar{Z}) \;. 
\end{array} \right. 
\label{2jastrow}
\end{equation}
The $z=0$ boundary condition $f^\prime(0^+)-f^\prime(0^-)=2f(0)/|a_{1D}|$, which 
accounts for the $\delta$-function potential, fixes the parameter $k$ 
through the condition $k|a_{1D}|\tan(kB)=1$. The remaining parameters $A$, $B$ and
$\beta$ are fixed by requiring that the function $f(z)$, its derivative 
$f^\prime(z)$ and the local energy $-2f^{\prime\prime}(z)/f(z)$ be continous 
at the matching point $z=\bar{Z}$. The value of the matching point $\bar{Z}$ 
is a variational parameter that we optimize using variational Monte Carlo 
(VMC). For $|z|<\bar{Z}$ the Jastrow 
term corresponds to the exact solution of the two-body problem and is expected
to provide a correct description of short-range correlations. Long-range 
correlations arising from phonon excitations are instead accounted for by the
functional dependence of $f(z)$ for $|z|>\bar{Z}$ \cite{Reatto}. Notice that
at $|z|=L/2$, which sets the limits of our simulation box, $f=1$ and particles 
are uncorrelated. The TG wave function $\Psi_0^{TG}=\prod_{i<j}
|\sin[\pi(z_i-z_j)/L]|$ is obtained as a special case of our trial function 
$\psi_T$ for $\bar{Z}=B=L/2$ and $kL=\pi$. The optimized Jastrow function introduced 
above gives for all values of $n|a_{1D}|$ variational results for the energy which
are almost identical to the exact DMC ones. The level of accuracy of the trial wave
function is particularly important for the calculation of $g_1(z)$. In fact, the pair 
distribution function $g_2(z)$ is calculated using the method of ``pure" estimators, 
unbiased by the choice of the trial function \cite{BORO2}. Due to the non-local 
property of the one-body density matrix, the function $g_1(z)$ can instead be obtained
only through the extrapolation technique. For an operator $A$ which does not 
commute with the Hamiltonian, the output of the DMC method is the mixed estimator 
$\langle\Psi_0|A|\psi_T\rangle$ and one can approximate the ``pure" estimator by
$\langle\Psi_0|A|\Psi_0\rangle = 2\langle\Psi_0|A|\psi_T\rangle - 
\langle\psi_T|A|\psi_T\rangle$, where $\langle\psi_T|A|\psi_T\rangle$ is the variational 
estimator. Of course, this procedure is very accurate only if $\psi_T\simeq\Psi_0$. 
We find 
that DMC and VMC give results for $g_1(z)$ which are very close and we believe that
the extrapolation technique is in this case exact. We also find that $g_1(z)$ is 
strongly affected by finite size effects particularly in the GP regime, where 
simulations of quite large systems are needed to obtain stable results. In order 
to control these effects we perform simulations with number of particles ranging 
from $N=100$ to $N=500$.

In Fig.1 we show the results of the pair distribution function for values of
$n|a_{1D}|$ ranging from the TG to the GP regime (the corresponding values of
the energy per particle are shown in the inset). For the smallest value of the 
gas parameter ($n|a_{1D}|=10^{-3}$) we find that $g_2(z)$ exactly coincides with
the TG result (\ref{g2tonks}). We also find that the values of 
$g_2(0)$ exactly reproduce the results obtained from the equation of state.
We notice that for small values of the gas parameter correlations die out very
quickly and $g_2(z)\simeq 1$ after few interparticle distances. For larger values 
of $n|a_{1D}|$ correlations extend over much larger distances. This result is also 
visible from the behavior of the static structure factor (shown in Fig. 2) which in 
the GP regime is dominated by long-wavelength fluctuations.

\begin{figure}
\begin{center}
\includegraphics*[angle=-90,width=0.95\columnwidth]{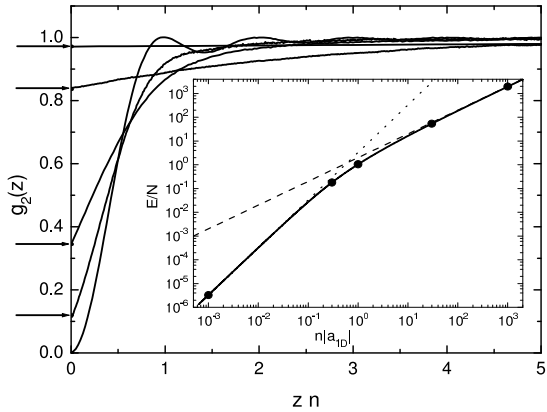}
\vspace{3 mm}
\caption{Pair distribution function for different values of the gas parameter. 
In ascending order of the value at $z=0$: $n|a_{1D}|=10^{-3}$, 0.3, 1, 30, $10^3$. 
Arrows indicate the value of $g_2(0)$ as obtained from the equation of state. 
Inset: energy per particle for the same values of $n|a_{1D}|$ (symbols), 
Lieb-Liniger equation of state (solid line), GP limit (dashed line), TG limit 
(dotted line). Energies are in units of $\hbar^2/(2ma_{1D}^2)$.}
\label{Fig1}
\end{center}
\end{figure}

In Fig. 3 we present results for the one-body density matrix. We see that for 
distances considerably larger than the healing length $\xi$, $g_1(z)$ is well 
reproduced by the asymptotic behavior (\ref{longOBDM}) for which we calculate 
the proportionality coefficient through a best fit to the DMC results. The 
deviations from the power law decay at the largest distances ($z\simeq L/2$) 
are due to finite size effects. In Fig. 4 we show the corresponding results 
for the momentum distribution. This is obtained from the Fourier transform 
of the calculated one-body correlation function at short distances and of 
the fits to the power-law decay (\ref{longOBDM}) at large distances. 
We notice that the infrared asymptotic behavior (\ref{nofkinfrared}) is 
recovered for values of $k$ considerably smaller than the inverse healing 
length $1/\xi$. At large $k$ the numerical noise of our results is too large 
to extract evidences of the $1/k^4$ behavior predicted in \cite{OLS-DUN}.

Let us now turn to harmonically trapped systems. In the 1D regime the system is 
described by the Hamiltonian $H_{LL}+\sum_{i=1}^{N}m\omega_z^2z_i^2/2$ with an
effective coupling constant given by $g_{1D}=2\hbar^2a/(ma_\perp^2)$, where $a$ 
is the 3D $s$-wave scattering length and $a_\perp=\sqrt{\hbar/(m\omega_\perp)}$ 
is the length fixed by the tight transverse confinement \cite{Olshanii,US}.
Consequently, the value of the effective 1D scattering length is given by the 
ratio $|a_{1D}|=a_\perp^2/a$. The relevant parameters are: the ratio $a/a_\perp$,
the anisotropy parameter $\lambda=\omega_z/\omega_\perp$ and the number of 
trapped particles $N$.

\begin{figure}
\begin{center}
\includegraphics*[angle=-90,width=0.95\columnwidth]{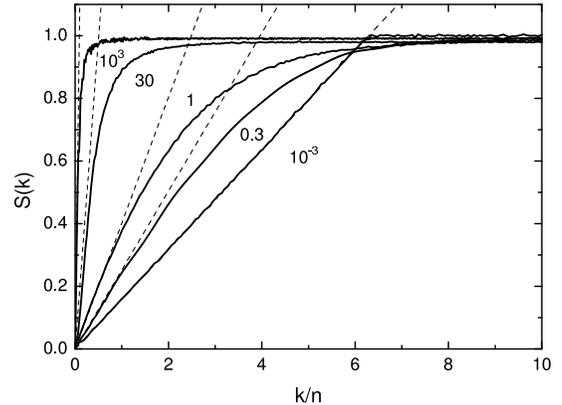}
\vspace{3 mm}
\caption{Static structure factor for the same values of $n|a_{1D}|$ as in Fig. 1 
(solid lines). The dashed lines are the corresponding long-wavelength asymptotics
from Eq. (\ref{sofkinfrared}).}
\label{Fig2}
\end{center}
\end{figure}

\begin{figure}
\begin{center}
\includegraphics*[angle=-90,width=0.95\columnwidth]{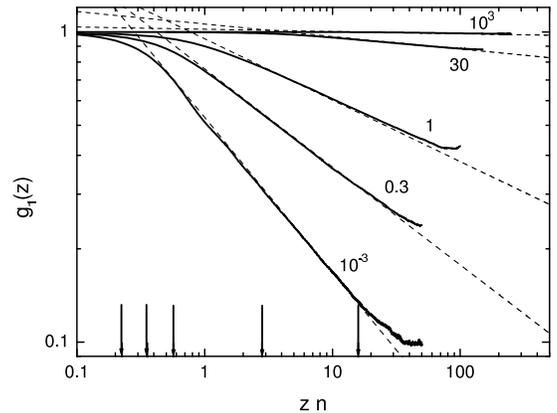}
\vspace{3 mm}
\caption{One-body density matrix for the same values of $n|a_{1D}|$ as in Fig. 1 
(solid lines). The dashed lines are the fits to the corresponding long-wavelength 
asymptotics from Eq. (\ref{longOBDM}). The arrows indicate the value of $\xi n$: 
the leftmost corresponds to $n|a_{1D}|=10^{-3}$, the rightmost to $n|a_{1D}|=10^{3}$.}
\label{Fig3}
\end{center}
\end{figure}

The DMC simulation is carried out using the following
trial wave function for the importance sampling: $\psi_T(z_1,..,z_N)=\prod_ig(z_i)
\prod_{i<j}f(z_{ij})$, where $g(z)=\exp(-\alpha_zz^2)$ is a one-body term which 
accounts for the confinement in the longitudinal direction and is fixed by the 
variational parameter $\alpha_z$. In the two-body term we neglect long-range 
correlations, which are irrelevant in the trapped case, and we choose $f(z)$ as 
given by (\ref{2jastrow}) for $|z|<\bar{Z}$ and $f(z)=1$ for $|z|>\bar{Z}$. 
We consider the following configurations: 
$a/a_\perp=0.2$, $\lambda=10^{-3}$ and number of particles  $N=$5, 20, 100.  
In Ref. \cite{US} we have proven that in these conditions the ground-state 
energy and structure of the cloud is correctly described by the Lieb-Liniger equation 
of state in local density approximation. The momentum distribution of a trapped system 
is obtained as: 
$n(k)=\int dZ dz^\prime \; n(Z) g_1(Z+z^\prime/2,Z-z^\prime/2) e^{ikz^\prime}$,
where $n(z)$ is the density profile and $n[(z+z^\prime)/2]g_1(z,z^\prime)= N \int 
\Psi_0^\ast(z,..,z_N)\Psi_0(z^\prime,..,z_N) dz_2 . . dz_N / \int 
|\Psi_0(z_1,..,z_N)|^2 \\ dz_1 . . dz_N$.

\begin{figure}
\begin{center}
\includegraphics*[angle=-90,width=0.95\columnwidth]{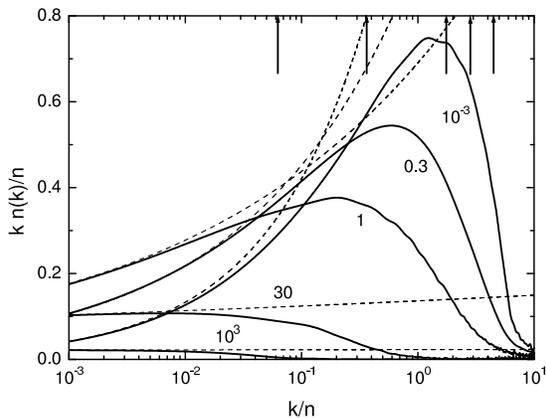}
\vspace{3 mm}
\caption{Momentum distribution for the same values of $n|a_{1D}|$ as in Fig. 1 
(solid lines). The dashed lines correspond to the infrared 
behavior of Eq. (\ref{nofkinfrared}). The arrows indicate the value of $1/(\xi n)$: 
the rightmost corresponds to $n|a_{1D}|=10^{-3}$, the leftmost to $n|a_{1D}|=10^{3}$.}
\label{Fig4}
\end{center}
\end{figure}

In Fig. 5 we show the results for the momentum distribution. Note the peaks becoming 
sharper with increasing $N$. In the case of $N=5$ and $N=20$, we find that the rounding 
off of $n(k)$ at $k\sim 1/R_z$, where $R_z$ is the size of the cloud in the axial 
direction, washes out completely the divergent behavior at small momenta. For the largest 
system with $N=100$ the value of the gas parameter is $n_0|a_{1D}|\simeq 1.1$, where $n_0$ 
is the density in the center of the trap, and the corresponding value of the exponent 
$\alpha$ is $\alpha\simeq 0.19$. In the region of wave vectors $1/R_z < k < 1/\xi$, where 
the healing length $\xi$ is estimated in the center of the trap, we find some evidence of 
the infrared behavior (\ref{nofkinfrared}) (see inset of Fig. 5). In order to see a cleaner
signature of this effect one should consider much larger systems.

In conclusion, we have carried out a complete study of the one- and two-body correlation 
functions of the 1D Lieb-Liniger model at $T=0$. We have also investigated how the presence 
of harmonic trapping modifies the behavior of the momentum distribution in configurations 
relevant to experiments.  

We would like to thank J. Boronat for useful discussions. This research is supported by 
the Ministero dell'Istruzione, dell'Universit\`a e della Ricerca (MIUR).

\begin{figure}
\begin{center}
\includegraphics*[angle=-90,width=0.95\columnwidth]{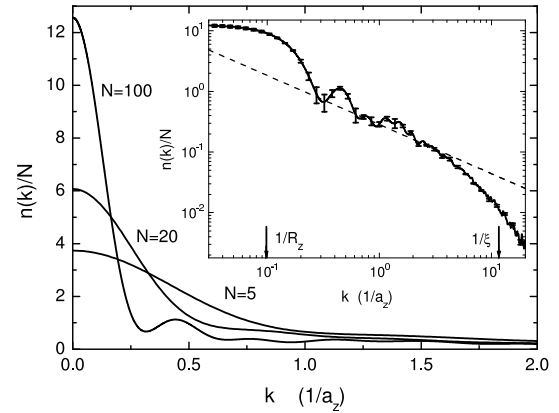}
\vspace{3 mm}
\caption{Momentum distribution of a trapped system. Inset: momentum distribution for 
$N=100$ (solid line) on a log-log scale. The dashed line is a fit to $1/k^{1-\alpha}$ 
with $\alpha=0.19$. The momentum distribution is in units of $a_z=\sqrt{\hbar/(m\omega_z)}$.}
\label{Fig5}
\end{center}
\end{figure}

\end{document}